
\magnification = \magstep1
\def\ltorder{\mathrel{\raise.3ex\hbox{$<$}\mkern-14mu
             \lower0.6ex\hbox{$\sim$}}}
\def\gtorder{\mathrel{\raise.3ex\hbox{$>$}\mkern-14mu
             \lower0.6ex\hbox{$\sim$}}}
\baselineskip = 18truept
\tolerance 10000
\def\ds{\displaystyle}
\centerline{\bf Pre-collapse Evolution of Galactic Globular Clusters}
\bigskip
\centerline{Toshiyuki Fukushige}
\smallskip
\centerline{Department of Earth Science and Astronomy,}
\centerline{College of Arts and Sciences,}
\centerline{University of Tokyo,}
\centerline{Komaba,}
\centerline{Meguro-ku,}
\centerline{Tokyo 153,}
\centerline{Japan}
\smallskip
\centerline{Email: fukushig@jp.ac.u-tokyo.c.chianti}
\medskip
\medskip
\centerline{Douglas C. Heggie}
\smallskip
\centerline{Department of Mathematics and Statistics,}
\centerline{University of Edinburgh,}
\centerline{King's Buildings,}
\centerline{Edinburgh EH9 3JZ,}
\centerline{UK}
\smallskip
\centerline{Email: d.c.heggie@ed.ac.uk}

\vfill\eject
{\bf Abstract}

This paper is concerned with collisionless aspects of the early evolution
of model star clusters.  The effects of mass loss through stellar
evolution and of a steady tidal field are modelled using $N$-body
simulations.  Our results (which depend on the assumed initial structure
and the mass spectrum) agree qualitatively with those of Chernoff \&
Weinberg (1990), who used a Fokker-Planck model with a spherically
symmetric tidal cutoff.  For those systems which are disrupted, the
lifetime to disruption generally exceeds that found by Chernoff \&
Weinberg, sometimes by as much as an order of magnitude.  Because we do
not model collisional effects correctly we cannot establish the fate of
the survivors.  In terms of theoretical interpretation, we find that
tidal disruption must be understood as a loss of {\sl equilibrium}, and
not a loss of {\sl stability}, as is sometimes stated.

{\bf Key words:} galaxies: star clusters -- globular clusters: general

\vfill\eject
\noindent
{\bf 1. Introduction}
\medskip
Our understanding of the evolution of an idealised globular cluster,
i.e.  an isolated, bound, spherically symmetric stellar system with
stars of equal mass, has increased rapidly in recent decades (Spitzer
1987, Djorgovski \& Meylan 1993).  Recently, the focus of interest has
moved to the study of the evolution of more realistic models of globular
clusters.  For example, stars in a real globular cluster have different
masses, and they change mass in response to their internal evolution.
Moreover, most globular clusters exist within galaxies, and are
influenced by the effects of the galactic tide.

These processes are linked.  Each star (above the turnoff mass
corresponding to the age of the system) loses a certain fraction of
its mass near the end of its own evolution, and this process decreases
the total mass of the cluster.  Therefore the potential of the cluster
weakens and the cluster expands.  As a result some stars flow over the
tidal boundary, and so further mass is lost by the cluster.  The time
scale of stellar evolution is $\sim 10^7$ years for a
$10M_{\odot}$ star, which is roughly comparable to the dynamical time
scale (or crossing time) of the cluster.  If the cluster contains a
large fraction of massive stars initially, the dynamics of the cluster
is greatly affected by the mass loss due to stellar evolution.

Chernoff and Weinberg (1990; hereafter CW) investigated these and other
aspects of the evolution of
globular clusters using Fokker-Planck models.  Their study
included the following realistic effects: the spectrum of stellar masses,
mass loss due to stellar evolution, and a tidal cutoff to model the
effect of the galactic tidal field.  They performed an extensive survey
of models differing with regard to the initial mass
function, the central potential of the cluster, and the galactocentric
distance.
For example,
they obtained the result that the mass loss during  $5\times10^9$ yr
is sufficiently strong to disrupt weakly bound clusters with a Salpeter
initial mass function.

The main purpose of the present paper is to check the results of CW
using a model which should be an improvement in several respects.  We
use direct $N$-body calculations, which allow us to include processes
taking place on the dynamical timescale.  These are neglected by CW, who
used an orbit-averaged method, and because the time scale for mass loss
and the dynamical time are not well separated, it is not clear {\sl a
priori} that their approximation is justified.  Like CW, we also include
the following ``realistic'' effects: a spectrum of stellar masses, mass
loss due to stellar evolution, and the tidal field of the galaxy.  This
last feature differs from CW's tidal cutoff, because it is not
spherically symmetric, and also because it affects stars even while they
remain inside the tidal radius.  Like CW, we performed a survey of King
models which differed in respect of the dimensionless central potential,
$W_0$, the slope ($\alpha$) of the initial stellar mass function
(assumed to be a power law), and the galactocentric distance, $R_{\rm
g}$, of the cluster.  In order to calculate the gravitational forces we
used a special purpose computer for the gravitational $N$-body problem:
GRAPE-3A (Okumura {\sl et al.}  1993).

In general we obtained qualitatively similar results to those of CW: the
less concentrated clusters ($W_0\ltorder4$) and/or ones that contained a
greater proportion of massive stars initially ($\alpha\gtorder -2.5$)
are disrupted before reaching the stage at which core collapse can
begin.  The main quantitative difference is that the lifetime to
disruption (for those clusters which do not survive until core collapse)
may be much longer than that found by CW.

The structure of the paper is as follows.  In section 2 we describe the
initial conditions and the way in which stellar mass loss and tidal
effects were modelled in our calculations.  (Most parameters were chosen
so as to be the same as those used by CW, for purposes of comparison).
In section 3 we describe briefly the hardware used for our calculations,
and the results are presented in \S4.  In section 5 we discuss the
mechanism of disruption, and \S6 sums up, with discussion of some
residual issues.

\bigskip
\noindent
{\bf 2. Model}
\medskip
\noindent
{\it 2.1 Equations of Motion}
\medskip
We adopt the conventional model in which the cluster is assumed to move
on a nearly circular orbit in a spherically symmetric galaxy potential,
taken to be that of a distant point mass $M_g$. Thus the stars are
affected by the steady tidal field of the galaxy, as well as by the
forces from other stars in the cluster.  We set the initial center of
mass of the cluster at the origin $(x,y,z)=(0,0,0)$, with axes
orientated so that the position of the galactic center is ($-R_g$,0,0).
We assume that the size of the globular cluster is much smaller than
$R_g$.  If the globular cluster rotates around the galactic center at an
angular velocity ${\bf \Omega} =(0,0,\omega)$ (cf.\S2.4), the equation
of motion of an individual cluster member can be expressed as $$
{d^2{\bf r}_i \over dt^2} = -\nabla\Phi_{\rm c,i} -2{\bf \Omega}\times
{d{\bf r}_i \over dt} +\omega^2(3x_i{\hat {\bf e}_x}-z_i{\hat {\bf
e}_z}), \eqno(1) $$ where ${\bf r}_i$ is the position of the $i$-th
star, and ${\hat {\bf e}_x},{\hat {\bf e}_z}$ are unit vectors that
point along the $x,z$ axes, respectively.  The first term on the
right-hand side in equation (1) is the gravitational acceleration from
other stars in the cluster, the second term is the Coriolis
acceleration, and the third term is a combination of the centrifugal and
tidal forces.  In our simulations we used a softened gravitational
potential, so that $$ -\nabla\Phi_{\rm c,i}= -\sum_{j=1,j\ne
i}^N{Gm_j(t)({\bf r}_i-{\bf r}_j)
\over (|{\bf r}_i-{\bf r}_j|^2+\varepsilon^2)^{3/2}},\eqno(2)
$$
where $N$ is the number of particles, $G$ is the gravitational
constant, and $\varepsilon$ is the softening parameter.

We performed numerical integrations of equation (1) using a second-order
predictor-corrector scheme with shared and constant timestep.  In models
with $N=16384$ the timestep, $\Delta t$, was 1/64, and the softening
parameter, $\varepsilon$, was 1/32, in standard units (Heggie \& Mathieu
1986), while for smaller values of $N$ we took $\Delta t=1/32$ and
$\varepsilon=1/16$. These values of $\varepsilon$ are comparable with
the mean interparticle distance.

\bigskip
\noindent
{\it 2.2 Initial Conditions}

\medskip
We used King's models (King 1966) to generate the initial conditions. Thus the
distribution function is a lowered Maxwellian, given by
$$
g(E)=K[e^{-\beta(E-E_t)}-1],\eqno(3)
$$
for $E<E_t=\phi(r_t)$, where $E=v^2/2+\phi(r)$, $K$ is constant, and
$r_t$ is the radius of the edge of the cluster.  (We use $\phi$ for
the Newtonian potential to distinguish it from the softened potential
$\Phi_c$.) The
King model is determined by the dimensionless central potential,
$W_0=\beta(\phi(r_t)-\phi(0))$.

We assign the masses of the stars according to a power-law mass
function: $$ dN(m)=m^{\alpha}dm,\eqno(4) $$ between $0.4M_\odot$ and
$15M_\odot$.  We assume that there is no correlation between the
position and mass; in particular there is no initial equipartition of
energies between stars of different mass.  Even at the present day mass
segregation is difficult to observe in the dynamically evolved globular
clusters of our galaxy, but our main reasons for making this assumption
are (i) simplicity and (ii) consistency with the assumptions of CW.

For this paper we performed a survey of models defined by combinations
of the values of the dimensionless central potential of the King model,
$W_0=1,3,5,7$, and the slope of the initial mass function,
$\alpha=-1.5,-2.5,-3.5$.  King models are not {\sl perfectly} suited to
the task, because the function $g(E)$ defined in eq.(3) is not an
equilibrium distribution for eq.(1), even if $\Phi_c$ is approximated by
a smooth potential, because $\phi$ excludes the tidal and centrifugal
potentials, and it is an unsoftened potential.  Therefore we first
carried out scaling of the stellar velocities to ensure virial
equilibrium, i.e. the condition $$ {\ddot {\sl I}}=0, \eqno(5) $$ where
$$ {\sl I}=\sum^N_{i=1}m_ir_i^2.  $$ This condition can also be
expressed as $$
\eqalignno{
{\ddot {\sl I}}&=\sum_{i=1}^N2m_i{\bf\dot r}_i^2 + \sum_{i=1}^N2m_i({\bf
r}_i\cdot{\ddot {\bf r}_i})\cr &=4T+2V_{\rm c}+V_{\rm s}-4V_{\rm g} =
0,&(6) } $$ where $$
\eqalignno{
T &= \sum_{i=1}^N{1\over2}m_i{\bf\dot r}_i^2,\cr
V_{\rm c} &=\sum_{i=1}^Nm_i{\bf r}_i\cdot(-\nabla\Phi_{{\rm c},i}), \cr
V_{\rm s} &=4\omega\sum_{i=1}^NL_{z,i}, &(7)\cr
\noalign{\hbox{{\rm and}}}\cr
V_{\rm g} &=\omega^2\sum_{i=1}^Nm_i({-{3\over 2}}x_i^2+{1\over 2}z_i^2).
} $$ Here, $T$ is the total kinetic energy, $L_{z,i}\equiv m_i(x_i\dot
y_i-\dot x_iy_i)$ is the $z$-component of the angular momentum of the
$i$th star, and $V_{\rm c}$, $V_{\rm s}$, and $V_{\rm g}$ are
contributions due to the gravitational force within the cluster, the
Coriolis force, and the combined centrifugal and tidal forces,
respectively.  For an unsoftened potential $V_c$ would be the potential
energy.  The external contribution to the energy of the cluster,
i.e. that due to tidal and centrifugal forces, is $V_g$.  The value of
$\omega$ was determined in such a way that the tidal radius (\S2.4)
coincided with the initial radius of the King model.

\bigskip
\noindent
{\it 2.3 Stellar evolution}

\medskip
We modelled the effect of stellar evolution by appropriately changing
the mass of each star.  At the last stage of stellar evolution stars
lose a significant fraction of their mass in winds and supernova
explosions.  The potential well of a cluster is not sufficiently deep to
retain the lost mass, since the escape velocity is only a few times
10km/s. Therefore we assume that the lost mass disappears abruptly from
the cluster.

We change the mass of each particle
according to
$$
m(t) =  \cases{\ds m_{\rm ini} &($t<t_{\rm se}$)\cr
		   m_{\rm rm} &($t\ge t_{\rm se}$)}\eqno(8)
$$
where $m_{\rm ini}$ is the initial mass, $m_{\rm rm}$ is the mass of any
remnant
after mass loss, and $t_{\rm se}$ is the time when the star
loses its mass (Table 1).  We
obtain values between the points listed in Table 1 by linear interpolation.
The remnant mass, $m_{\rm rm}$, is summarized in Table 2.
These tables are due to Iben and Renzini(1983), and were also those used
by CW, from which these tables have been copied.

\bigskip
\noindent
{\it 2.4 Tidal Boundary}
\medskip
During the course of a simulation stars escape from the cluster.  The
precise dynamical definition of escape is not easy if there is a tidal
field, and here we adopt a simple geometric definition: escapers are
defined to be those stars beyond the tidal radius.  All stars within the
tidal radius are taken to be members, even though
the tidal field included in equation (1) is not spherically
symmetric.  More precisely, we use the distance between the center of
the cluster (defined below) and the
Lagrangian point in the direction of the galactic center as the tidal
radius. If, as before, the galaxy is represented by a distant point
mass, it follows that the tidal radius is given by
$$
r_{\rm t} = \ds\left({M\over 3M_{\rm g}}\right)^{1/3}R_{\rm g},\eqno(9)
$$
approximately, where $M$ is the mass of the cluster, $M_{\rm g}$ is
the mass of the galaxy and $R_{\rm g}$
is the distance
to the galaxy. Here $M$ is taken to be the total mass of the `members',
and since this depends on $r_t$ itself, some iteration is usually
required.  Since the angular velocity of the cluster around the galaxy
is given by $\omega^2 = GM_g/R_g^3$, it follows that
$$
\omega = \ds\sqrt{GM\over 3r_{\rm t}^3}.\eqno(10)
$$

We define the (density weighted) center of the cluster by
$$
{\bf r}_{c} = \sum_{i=1}^N \rho_i^2{\bf r}_i / \sum_{i=1}^N \rho_i^2, \eqno(11)
$$
where $\rho_i$ is the local density around the $i$-th particle.  For
this purpose we use the
local density that was introduced by Casertano \& Hut (1985) and
defined by
 $$
\rho_i = {3M_{6,i}\over 4\pi R_{6,i}^3},\eqno(12)
$$
where $R_{6,i}$ is the distance to the sixth nearest neighbor of star
$i$, and
$M_6$ is the total mass lying within the distance $R_{6,i}$
from star $i$.

\bigskip
\noindent
{\it 2.5 Scaling}
\medskip
We use standard units such that $M=G=-4E=1$, where $M$ is the initial
total mass and $E$ is the initial total self-energy (Heggie and Mathieu
1986), and we now consider how the results should be scaled to
astrophysical units. Simple $N$-body simulations have two degrees of
freedom for scaling, e.g. the units of length and time.  However, we
have to specify the unit of time in years in order to determine the time
of mass loss by stellar evolution in the simulation.  Thus two
simulations in which the value of the unit of time differs will
experience different evolution in general.

We wish to select scalings appropriate to a cluster of mass $M$, at
galactocentric radius $R_{\rm g}$, moving at circular speed $v_{\rm g}$,
with mean stellar mass $\langle m\rangle$, and tidal radius $r_t$, so
that it may be simulated by a model composed of $N$ particles with
standard units.  Because of the definition of these units we select the
units of mass, $U_{\rm m}$, and length, $U_{\rm l}$, as follows; $$
\eqalignno{
U_{\rm m} &= M,&(13)\cr U_{\rm l} &= r_{\rm t}/r^{\ast}_{\rm t}&(14), }
$$ where $r^{\ast}_{\rm t}$ is the tidal radius of the King model in
standard $N$-body units.  Since a cluster with $M/\langle m \rangle$
stars is being modelled using $N$ particles, a particle in the
simulation of mass $m^{\ast}$ ($\sum m^{\ast}=1$) represents a total
mass $Mm^{\ast}$ which consists of $(M/\langle m \rangle)/N$ stars of
individual mass $Nm^{\ast}\langle m \rangle$.

Conversion to standard units also requires that  the unit of time is
given by
$$
U_t = \ds\left({U_{\rm l}^3\over GU_{\rm m}}\right)^{1/2}
= \left({r_{\rm t}^3\over GMr_{\rm t}^{\ast3}}\right)^{1/2}
= {R_{\rm g}\over v_{\rm g}}\left({1\over 3r_{\rm t}^{\ast3}}
\right)^{1\over 2}, \eqno(15)
$$ where we have used eqs.(9), (13) and (14).  It follows that the
crossing time of the $N$-body model scales correctly to that of the real
globular cluster.  Also, to determine when a given star loses mass, we
compute $Nm^{\ast}\langle m \rangle$, and obtain the evolution time,
$t_{\rm se}$, from Table 1.  Then, the corresponding $N$-body time is
$t_{\rm se}/U_{\rm t}$.

There are two characteristic time scales for the evolution of a globular
cluster: the crossing time, $t_{\rm cr}$, characterizing the orbital
motions of the stars, and the relaxation time, $t_{\rm r}$,
characterizing two body relaxation.  Ideally, we need to follow the
evolution of a globular cluster with a technique which correctly models
effects on both time scales.  The time scale of the internal evolution
of massive stars is comparable to the dynamical (crossing) time scale of
a typical globular cluster ($\ltorder 10^7$ years).  If the fraction of
massive stars is not very small, the globular cluster evolves
significantly within the dynamical timescale.  On the other hand,
relaxation is equally important in the long run. Evaporation due to
two-body encounters reduces the mass of the cluster, while relaxation
increases the central density, thus helping the cluster to avoid
disruption by the tidal field.

The ratio of the crossing time to the relaxation time (within the
half-mass radius) is approximately linearly dependent on the particle
number $N$ as follows (Spitzer 1987): $$ {t_{\rm r}\over t_{\rm cr}} =
{N\over 11\ln{(0.4N)}}.\eqno(16) $$ When we model a cluster consisting
of $\sim 10^5-10^6$ stars using a smaller number of particles (in most
of our simulations $N = 8192$), the ratio is different from that of the
real globular cluster system.

Under the scaling that we have adopted, the results of the simulation
have little significance after the time when the system is considerably
affected by two-body relaxation, since the system will become
relaxed much sooner than a real globular cluster.  We used a relatively  large
number of particles, $N$, and a large softening parameter, $\varepsilon$,
comparable to the mean particle separation, in order to suppress two body
relaxation effects as much as possible.

As an alternative, it would be possible to choose a scaling which
ensures that the {\sl relaxation} of the simulation takes place on the same
time scale as that of the real cluster (Giersz \& Heggie 1994; cf.
Heggie 1994).  On the other hand this implies that the crossing time of
the simulation is much greater than it should be, and so loss of mass by
stellar evolution is much more impulsive in character than in a real
cluster.

\bigskip
\noindent
{\bf 3. Hardware}
\medskip
For the force calculation, we used a special-purpose computer for
$N$-body problems, called GRAPE-3A, which is a modified version of
GRAPE-3 (Okumura {\sl et al.}  1993).  It consists of 8 LSI chips
dedicated to the calculation of gravity and has a peak performance of
about 5 Gflops-equivalent per board (at 20MHz clock cycle).  We also
used the nearest-neighbor list produced by GRAPE-3A to compute the local
density as defined in section 2.4.

GRAPE-3A is a special-purpose computer mainly for collisionless $N$-body
calculations.  When calculated with GRAPE-3A the force between two
particles has a relative error of a few percent, because of the low
accuracy of internal expressions in the GRAPE chip.  However, GRAPE-3A
can follow properly the evolution of self gravitating systems unless the
system contains a high-density core or its evolution is governed by the
behaviour of binaries.  According to Hernquist {\sl et al.}  (1993), the
change in velocity due to the error of GRAPE-3A, denoted by $v_{err}$,
may be estimated by $$
\langle v_{err}\rangle^2 \le e^2 \langle v_{2b}\rangle^2,\eqno(17)
$$ where $\langle v_{2b}\rangle$ is the change in velocity due to 2-body
relaxation effects.  Here, $e$ is the relative error of the force
between two particles.  In the case of GRAPE-3A, $e$ is estimated at
$2\sim 3\%$ (Makino {\sl et al.} 1990; Okumura et al.~1993).  Therefore,
the error due to GRAPE-3A is small compared to two-body relaxation
effects, and we have already pointed out that these are required to be
small if our simulations are to be valid.

\bigskip
\noindent
{\bf 4. Results}
\bigskip
\noindent
In this section we present the results of our calculations, in which we
performed a survey of models differing in the slope, $\alpha$, of the
initial power-law mass function, and in the dimensionless central
potential of King's model, $W_0$.  In section 4.1 we summarize the
results of the survey.  In section 4.2 we investigate the evolution of
the total mass for each set of parameters.  In sections 4.3 and 4.4 we
discuss the effects of varying the galactocentric distance and of
two-body relaxation, respectively.  In section 4.5 we present data on
the anisotropy of the velocity dispersion.

\bigskip
\noindent
{\it 4.1. Summary of Survey Results}
\medskip
In Table 3, the parameters of all runs are listed.  The choices of
$\alpha$ and $W_0$ were made for consistency with those of CW, except
that we added the cases in which the dimensionless central potential
$W_0 =5$.  Now the choice of $U_t$ will be explained.  CW defined four
families of models in terms of a combination of parameters which we
denote by $F_{cw}$, and which is defined by $$ F_{\rm cw} =
\ds\left({M\over M_{\odot}}\right) \left({R_{\rm g}\over Kpc}\right)
\left({220kms^{-1}\over v_{\rm g}}\right) \left({1\over \ln{(M/\langle m
\rangle)}}\right).\eqno(18) $$ In the case of family 1 of CW, this value
was given as $F_{\rm cw}=5.0\times 10^4$. For a tidally truncated
cluster with a given structure and mass spectrum, $F_{cw}$ is a measure
of the mean relaxation time, and different clusters within this family
have different crossing times.  In our $N$-body models, the evolution of
systems with different crossing times will differ (in principle), and so
we must choose a specific model from family 1 of CW for comparison.  CW
themselves gave as an example a cluster with total mass, $M$, equal to
$1.49\times 10^5M_\odot$, and took $v_g$, the circular speed around the
galaxy, to be $220kms^{-1}$.  If we specify the slope of the IMF and the
maximum and minimum stellar masses, the mean mass $\langle m \rangle $
is determined.  Then, for the stated value of $F_{\rm cw}$, we obtain
the galactocentric distance, $R_{\rm g}$, from equation (18), and the
unit of time is determined by equation (15).  In this way the
galactocentric distance, $R_{\rm g}$, is calculated as 3.7Kpc for
$\alpha=-1.5$, 4.0Kpc for $\alpha=-2.5$, and 4.1Kpc for $\alpha=-3.5$.
We list the corresponding unit of time $U_{\rm t}$ for each set of
parameters in Table 3.  Note, however, that our models equally well
represent the evolution of clusters within CW's other 3 families
(cf.\S4.3), for appropriate values of the various parameters, and
provided always that relaxation is negligible. The particle number, $N$,
in our survey was usually $N=8192$.

The lifetime of the cluster is also listed in Table 3 for each set of
parameters.  The first and second columns are the dimensionless central
potential, $W_0$, and the slope of the power-law IMF, $\alpha$.  The
third column is the unit of time calculated by means of equation (15).
In the fourth column, the lifetime of the cluster is given in $N$-body
units and in years (in brackets).  The letter S means that the cluster
survived at the end of calculation, which was set at $T=2000$ in
$N$-body units.  We determined this value in relation to the initial
half-mass relaxation time $t_{\rm rh}\sim 700$ for $N=8192$ and
$\varepsilon=1/16$.  The fifth column is the lifetime obtained by CW.
The letter C means that they found that the cluster experiences core
collapse.  The last column, which concerns anisotropy, is discussed in
\S4.5.

\bigskip
\noindent
{\it 4.2. Evolution of Total Mass}
\medskip
In this and the following subsections we discuss several aspects of our
models in some detail.
Figure 1 shows the evolution of the total mass, $M$, of the cluster.  The bold
curves indicate the decrease of mass of the cluster defined by those
stars within the
tidal boundary.  The thin curves indicate the
decrease of the mass of all stars due to stellar evolution only.

As can be seen from Table 3 and Fig.1, the less concentrated clusters
and/or those containing more massive stars are disrupted sooner.  When
the initial concentration of the cluster is small and it contains
relatively many massive stars, it disrupts within about $10\sim 20$
crossing times.  The runs for $(W_0,\alpha)=(1,-1.5)$ and $(3,-1.5)$ are
representative of this behaviour.  In these cases, massive stars evolve
immediately and the cluster loses a large fraction of its mass.  Since
the cluster is not at all concentrated, it is easily disrupted by the
tidal field of the galaxy. Details of the evolution of the spatial
structure are shown in figure 2, which displays Lagrangian radii for the
case $(W_0,\alpha)=(3,-1.5)$.  The number over each curve denotes the
fraction (in percent) of the {\sl initial} total mass.

When the cluster initially has fewer massive stars than in the above
cases, we can observe rapid disruption at the final stage of the
evolution.  $(W_0,\alpha)=(1,-2.5)$, $(1,-3.5)$ and $(3,-2.5)$ are
examples of this kind of evolution.  In these cases, before this happens
the cluster {\sl gradually} loses its mass by the following two
processes.  First, the mass loss by stellar evolution weakens the
potential well of the cluster and the cluster expands.  Then, the stars
outside the tidal boundary escape from the cluster.  Moreover, the
evaporation of stars due to two body relaxation decreases the mass of
the cluster, though we have taken care that this has a minor influence
on our results (cf.\S4.4).  After the cluster has reached a certain
critical point, it disrupts more-or-less immediately (cf.\S5 for a
detailed investigation of this final disruption process). In figure 3,
the Lagrangian radii are plotted for the case of
$(W_0,\alpha)=(3,-2.5)$.

When the cluster is initially concentrated, or does not contain many
massive stars, it does not disrupt (within the duration of the
simulation).  Then the cluster may experience core collapse. In figure
4, the Lagrangian radii are plotted for the case of
$(W_0,\alpha)=(3,-3.5)$, which is typical of this behaviour.

\bigskip
\noindent
{\it 4.3. Effect of galactocentric distance}
\medskip
Figure 5 shows the evolution of the total cluster mass for clusters of
the same initial mass at four different galactocentric distances,
$R_{\rm g}$, in the case of $W_0=3$ and $\alpha=-2.5$.  The number of
particles is always $N=8192$.  In the calculations, the different
$R_{\rm g}$ correspond to different values of the unit of time, $U_t$,
according to equation (15).  The numbers in the Figure indicate the
galactocentric distance if, as before, we set the mass of the cluster to
be $M=1.49\times 10^5M_{\odot}$.  These four galactocentric distances,
4.0, 10.6, 18.0 and 47.4 kpc, correspond to the families 1, 2, 3 and 4
of CW, respectively. In table 4, the lifetimes of the clusters are
summarized.

\medskip

\bigskip
\noindent
{\it 4.4. Effect of two-body relaxation}
\medskip
We investigated the effect of two-body relaxation by means of calculations
with different particle number, $N$.  Figures 6, 7, and 8 show the evolution
of the total cluster mass for several different particle numbers in the cases
$(W_0,\alpha)=(3,-1.5)$, $(3,-2.5)$ and $(1,-3.5)$, respectively.
The calculations were performed with $N=1024$, 2048, 4096 and 8192 for all
cases and $N=16384$ for $(W_0,\alpha)=(3,-2.5)$ and $(1,-3.5)$.

As shown in figure 6, the cluster disrupts in several crossing times in
the case of $(W_0,\alpha)=(3,-1.5)$.  Different values of the particle
number, $N$, do not significantly affect the result.  In this case, the
cluster disrupts only through processes occurring on the dynamical and
stellar evolution time scales.

As shown in figure 8, the cluster disrupts after approximately 300
crossing times in the case $(W_0,\alpha)=(1,-3.5)$.  The result depends
greatly on the number of particles $N$, until this is rather large.  In
this case, unless $N$ is very large the cluster loses mass due to both
stellar evolution and evaporation by two-body relaxation.  When the
cluster loses enough mass and reaches a certain critical structure, it
disrupts immediately.  Clusters with smaller particle number lose more
mass, because the two-body evaporation effect is stronger.  Moreover,
those clusters with the smallest particle numbers ($N\le 4096$) would
experience core collapse well within the duration of these simulations
(if the evolution were followed with an accurate, collisional code), and
this would help to prevent tidal disruption. Nevertheless the similarity
of the two largest simulations in Fig.8 suggests that relaxation effects
are relatively unimportant for them.  Assuming that mass loss from
evaporation in a given time varies approximately as $N^{-1}$ (i.e. in
proportion to the reciprocal relaxation time), one may estimate that the
lifetime to disruption of the model with $N=16384$ differs from the
result for very large $N$ by about 10\% only.

As shown in figure 7, the case $(W_0,\alpha)=(3,-2.5)$ is intermediate
between the above two cases. The clusters with smallest particle number
($N\le 2048$) are substantially affected by two-body relaxation, and
when $N=1024$ it survives without disruption.  The results with large
particle number ($N\ge 8192$) differ very little from each other.  This
shows that the decrease of mass is determined almost entirely by stellar
mass loss.

\bigskip
\noindent
{\it 4.5. Anisotropy in Velocity Dispersion}
\bigskip
\noindent
Figure 9 shows data on the anisotropy in the velocity distribution of
the particles.  Here the anisotropy is measured by the parameter $$ A =
2- {\langle v_{\rm t}^2 \rangle \over \langle v_{\rm r}^2 \rangle
},\eqno(19) $$ where $\langle v_{\rm t}^2 \rangle$ and $\langle v_{\rm
r}^2 \rangle$ are the mean square tangential and radial velocities,
respectively, of particles within the tidal radius.  The approximate
minimum value of $A$ throughout the lifetime of each model is given in
the last column of Table 3.

The parameter $A$ vanishes for an isotropic distribution of velocities
(as is true initially), and would be positive if there were a
preponderance of radial orbits, as happens in the evolution of isolated
models driven by two-body relaxation only (cf. Spitzer 1987). In all
cases which we have computed, however, there is a tendency for $A$ to
decrease, which implies a deficit of stars on nearly radial orbits.  We
interpret this as being due to the preferential escape of stars moving
on radial orbits.  In those clusters which disrupt within the time scale
of our simulations, however, the disruption is signalled by a rather
sharp rise in $A$.  This presumably arises from the predominantly radial
motions of the stars as the cluster finally dissolves.

Figure 10 shows one component of the angular velocity, $\Omega_{\rm
c}$, of the  case $W_0=3$, $\alpha=-2.5$. This is defined here by
$$
\Omega_{\rm c} = {L_z\over \sum m_i(r_{i,x}^2+r_{i,y}^2)},\eqno(20)
$$ where $L_z$ is the $z$-component of angular momentum, and $r_x$ and
$r_y$ are the $x$- and $y$-coordinates of the $i$th star relative to the
density center.  The angular velocity is so small that the rotation of
the cluster is insignificant dynamically; the corresponding term $V_s$
in the virial relation, eq.(6), is almost always negligible.  The
initial decrease in $\Omega_c$ may be interpreted as being due to the
expansion of the cluster following mass loss within the first few
crossing times: the Coriolis term in eq.(1) causes $L_z$ to decrease if
the radial motion is predominantly outwards.

\bigskip
\noindent
{\bf 5 Mechanism of Disruption}
\medskip
In this section, we investigate the mechanism of disruption of a
cluster.  As shown in figure 1, the final stages of this process can occur
rapidly, which suggests that disruption could be
caused by a loss of equilibrium.  The
abruptness with which disruption occurs is shown even more clearly in Figure
11, which plots
the virial ratio, defined by
$$
q = -4T/(2V_{\rm c}+V_{\rm s}-4V_{\rm g})\eqno(21)
$$
(cf. eq.(6)) for four of those clusters which exhibit
disruption.  This figure shows that when rapid disruption
occurs the cluster loses virial balance.

When the cluster is in dynamical equilibrium, as in eq.(6) we have
$$
{\ddot {\sl I}} = 4T+2V_{\rm c}+V_{\rm s}-4V_{\rm g} = 0.\eqno(22)
$$
If we eliminate $T$ using the total energy of the cluster, $E$, given by $E = T
+ V_{\rm c} + V_{\rm g}$,
we obtain the relation:
$$
{\ddot {\sl I}} = 4E-2V_{\rm c}+V_{\rm s}-8V_{\rm g}.\eqno(23)
$$
Therefore the equilibrium condition is expressed as
$$
4E = 2V_{\rm c}-V_{\rm s}+8V_{\rm g}.\eqno(24)
$$
To analyse this condition we reexpress $V_{\rm c}$ and $V_{\rm g}$ as
$$
\eqalignno{
V_{\rm c} &= -{\mu GM^2 \over r_{\rm h}},&(25)\cr
V_{\rm g} &= -{\nu \omega^2Mr_{\rm h}^2},&(26)
}
$$
where $\mu$ and $\nu$ are dimensionless parameters, and
$r_{\rm h}$ is the half mass radius of the cluster.
If we substitute equations (25) and (26) into equation (24), this condition
becomes
$$
4E+V_{\rm s} = -{2\mu GM^2 \over r_{\rm h}}
-{8\nu \omega^2Mr_{\rm h}^2}.\eqno(27)
$$
Finally, by eliminating $\omega$ in favour of
the tidal radius $r_{\rm t}$, the equilibrium condition
is expressed as
$$
{r_{\rm t}(4E+V_{\rm s})\over 2GM^2}
= -{\mu \over (r_{\rm h}/r_{\rm t})}
-{4\nu\over 3}\left({r_{\rm h}\over r_{\rm t}}\right)^2.  \eqno(28)
$$

When $r_{\rm h}/r_{\rm t}$ is small the right-hand side of equation (28)
is dominated by the first term, which comes from the self-potential of
the cluster, while when $r_{\rm h}/r_{\rm t}$ is large it is dominated
by the second term, which is due to the potential of the galaxy and the
centrifugal force.  Therefore, the right-hand side of equation (28),
when regarded as a function of $r_h/r_t$, has a maximum value at a
particular point, if $\mu$ and $\nu$ are constant (cf. Fig.13).  (In
practice these parameters are indeed nearly constant, as shown in one
case below.)  Initially the cluster has a value of $r_h/r_t$ smaller
than that at the maximum point, and the value of the function on the
left side of eq.(28) is below the maximum value.  As the cluster loses
mass this quantity increases, as does the ratio $r_{\rm h}/r_{\rm t}$,
and the cluster moves towards the maximum point along the curve.  When
the function on the left side of eq.(28) exceeds the maximum value on
the stated curve there is no longer a solution on the equilibrium curve.
Therefore the cluster loses equilibrium and disrupts.

It is not hard to understand why the ratio $r_h/r_t$ increases, but the
reason for this behaviour is not simply the expansion resulting from
mass loss.  As the cluster loses mass due to stellar evolution, it does
indeed expand, and the half mass radius increases.  But the particles
which move beyond the tidal radius cease to be members, and the tidal
radius decreases in proportion to $M^{1/3}$.  There is a consequent {\sl
reduction} in $r_h$.  However, since the density around the tidal radius
is smaller than that around the half mass radius, $r_h$ does not
decrease so much.  Therefore, the ratio $r_{\rm h}/r_{\rm t}$ becomes
larger.

We illustrate this loss of equilibrium using numerical data for the case
$W_0=1$, $\alpha=-3.5$ and $N=16384$.  Figure 12 confirms that the
parameters $\mu$ and $\nu$ are almost constant until the cluster loses
equilibrium.  We plot the equilibrium curve, eq.(28), using constant
values $\mu=0.42$ and $\nu=0.52$, which gives the bold curve in figure
13.  The star-shaped symbols plotted in the figure were obtained from a
simulation, for which data are plotted every 100 time units from $t=0$
to $t=900$, and every 10 time units after $t=900$.  As the cluster loses
mass, the point representing the cluster moves upward and to the right
on this curve, while the cluster maintains virial equilibrium.  When the
dimensionless energy becomes larger than the maximum value on the
equilibrium curve, the cluster loses virial equilibrium, and the
disruption is extremely rapid.

\bigskip
\noindent
{\bf 6. Discussion}
\bigskip

Our $N$-body calculations have been designed to investigate the
survival of large star clusters (i.e. those with large $N$, such as
globular star clusters) against two important disruptive processes:
(i) mass loss resulting from the internal evolution of the stars, and
(ii) tidal stripping due to the gravitational field of a parent
galaxy.  The importance of the size of $N$ stems from the fact that we
have suppressed two-body relaxation effects as much as possible, and
so our models are only applicable to large star clusters in which
relaxation effects are negligible over timescales equivalent to the
duration of our simulations.  We checked that relaxation is negligible
in our models by repeating runs, especially the longest ones, with
much larger values of $N$, and also (though this was not discussed in
our paper) by checking for mass segregation.  Indeed, one way of
suppressing relaxation even further in our models would have been to
use models with particles of equal mass, each of which changes with
time in proportion to the mass of the entire population.
Conceptually it was simpler to use a distribution of particle masses
and to let each particle lose mass at an appropriate time, but this
did have the result of enhancing two-body effects through mass
segregation.

Our model of mass loss through stellar evolution is a simplification
of what is thought to happen in nature.  In particular, we assume that
each star abruptly loses a certain fraction of its mass at the end of
its internal evolution, the time at which this occurs, and the
fraction of mass, being determined uniquely by its initial mass.  No
attempt was made to model gentler phases of mass loss, via stellar
winds, for example.  Though the assumptions are a considerable
idealisation, they were partly made for consistency with the work
of Chernoff \& Weinberg (1990), as one aim of our models was to verify
their results with a different dynamical model.

As with our  treatment of stellar evolution, we have modelled the
galactic tide in a very simplified but conventional manner, i.e. as
that due to a distant point mass, about which the cluster describes a
circular orbit.  In particular this neglects any time-dependence in
the tide.  The time-dependence of the  tide is almost certainly of
importance for an adequate understanding of the history of the
globular clusters which survive to the present day.  On the other hand a
steady tide illustrates some important aspects of tidal stripping, and
we have improved the treatment of Chernoff \& Weinberg by including
the acceleration of the tide in the equations of motion of all stars,
and not simply as a cutoff which demarcates the bound from the unbound
stars.  In addition our model of the tide is not spherically
symmetric, unlike the treatment of Chernoff \& Weinberg.

The evolution of a typical model may be summarized as follows.  Mass
loss due to stellar evolution decreases the total mass of the cluster.  In
consequence the tidal radius decreases, and the stars outside the
tidal radius are stripped off.  The mass loss from stellar evolution
also weakens the gravitational potential.  One consequence is that the
cluster expands, and so some stars overflow the tidal boundary.
(Also, the total mass of the cluster decreases as a result of evaporation
caused by two-body encounters, but, as already explained, we have
checked that this effect is relatively weak.)  Though mass loss by
both effects (stellar evolution and tidal stripping) always occur
simultaneously, as the cluster loses mass the tidal effect becomes
progressively stronger relative to the effect of stellar evolution.
When the influence of the
tide on the virial balance of the cluster becomes sufficiently large
relative to its
self-gravitation, the whole cluster loses dynamical
equilibrium, and disrupts rapidly.  In some of our models, however,
especially those in which  the rate of mass loss due to stellar
evolution is small, the cluster does not reach this critical point
within the duration of our simulations (which is determined by the
onset of two-body relaxation effects).  In these cases the ultimate
fate of the cluster  cannot be determined from our models, though it
seems likely that such systems would eventually evolve towards core collapse.

As has been stated, one of the main purposes of our investigation was
to verify the results of Chernoff \& Weinberg (1990), which have
proved such an important and stimulating source of information on the
dynamical evolution of tidally bound clusters which lose mass by
stellar evolution.  Unlike us, they did include two-body relaxation
effects.  In other respects their models were somewhat more idealised;
besides the treatment of the tide, to which we have already referred,
their models treated the distribution of stellar velocities as
isotropic.

In view of these differences it is interesting to find that our
results do not differ qualitatively from those of Chernoff \&
Weinberg:  the models which disrupt tidally are the same in both
studies, and the models which survive until core collapse (according
to CW) are those which, in our study, avoid tidal disruption to the
point at which two-body relaxation may become effective.  Where our
results differ is in the time to disruption, for those models which
disrupt tidally.  Our models almost always outlast those of CW, and sometimes
by a factor of ten or more.  It is impossible to determine which of
the differences in the models is responsible for this disagreement.  In
addition to the aspects discussed above, our models are able to follow
processes occurring on a dynamical (crossing) time scale, and so there
are several factors which could contribute to this result.

It is clear, however, that the models we have studied are not
necessarily the most important for future cluster studies.  In
particular, the lower limit of the initial mass function is almost
certainly too high, and the addition of an appropriate number of low
mass stars would enhance the survivability of the clusters.  In
addition, non-steady tides (e.g. disk-shocking) are an important
mechanism, missing in both studies, which acts in the opposite
direction.  Our results do, however, suggest that a Fokker-Planck
model with a spherically symmetric cutoff and an isotropic
distribution of velocities can be quantitatively very unreliable.

\bigskip
\noindent
{\bf Acknowledgements}

The GRAPE hardware at Edinburgh was acquired using a grant (No.GR/H93941) from
the UK Science and Engineering Research Council.  The computations for
this paper were carried out by TF while visiting Edinburgh with the
financial support of SERC (Grant No.GR/H95303) and the British Council
(Tokyo), to both of whom we express our thanks.  We are grateful to
Mirek Giersz (supported by SERC Grant No.GR/G04820) for much help in setting
up the initial conditions of our models.  TF thanks Daiichiro Sugimoto
and Junichiro Makino for helpful discussions.  This work was carried out
while TF was a fellow of the Japan Society for the Promotion of Science
for Japanese Junior Scientists.

\bigskip
\noindent
{\bf References}

{\leftskip=20pt\parindent -20pt
\bigskip
Casertano S., Hut P., 1985, ApJ, 298, 80

Chernoff D.F., Weinberg M.D., 1990, ApJ, 351, 121

Djorgovski S.J., Meylan G., eds, 1993, Structure and Dynamics of
Globular Clusters, ASP Conf. Ser., Vol 50. ASP, San Francisco.

Giersz M., Heggie D.C., 1994, in preparation

Heggie D.C., 1994, in W. Wamsteker, M.S. Longair, Y. Kondo, eds,
Frontiers of Space and Ground-Based Astronomy: The Astrophysics of the
21st Century, Proc. 27th ESLAB Symposium. Kluwer, Dordrecht, in press

Heggie D.C., Mathieu R.D., 1986, in Hut P., McMillan S., eds, The Use
of Supercomputers in Stellar Dynamics, LNP 267. Springer-Verlag, Berlin, p.233

Hernquist L., Hut P., Makino J.,  1993, ApJ, 402, L85

Iben I.J., Renzini A., 1983, ARA\&A, 21, 271

King I.R., 1966, AJ, 71, 64

Makino J., Ito T., Ebisuzaki T., 1990, PASJ, 42, 717

Okumura S. K., Makino J., Ebisuzaki T., Fukushige T., Ito T., Sugimoto D.,
Hashimoto E., Tomida K., Miyakawa N.,  1993, PASJ, 45, 329

Spitzer L., Jr., 1987, Dynamical Evolution of Globular Clusters.
Princeton U.P., Princeton.

}

\vfill\eject
\centerline{Table 1. Stellar evolution time}
$$\vbox{
\halign{\hfil#\hfil&\quad\hfil#\hfil\cr
\noalign{\hrule}
\noalign{\vskip 2 pt}
\noalign{\hrule}
\noalign{\vskip 5 pt}
Initial Mass $m$ & Main Sequence Time \cr
$(\log_{10}[m/M_{\odot}])$ & $(\log_{10}[t_{\rm se}/yr])$ \cr
\noalign{\vskip 5 pt}
\noalign{\hrule}
\noalign{\vskip 5 pt}
1.79 & 6.50 \cr
1.55 & 6.57 \cr
1.33 & 6.76 \cr
1.11 & 7.02 \cr
0.91 & 7.33 \cr
0.72 & 7.68 \cr
0.54 & 8.11 \cr
0.40 & 8.50 \cr
0.27 & 8.90 \cr
0.16 & 9.28 \cr
0.07 & 9.63 \cr
$-$0.01 & 9.93 \cr
$-$0.08 & 10.18 \cr
\noalign{\vskip 5 pt}
\noalign{\hrule}
}
}$$
\bigskip
\centerline{Table 2. Mass evolution}
$$\vbox{
\halign{\hfil#\hfil&\quad\hfil#\hfil&\quad\hfil#\hfil\cr
\noalign{\hrule}
\noalign{\vskip 2 pt}
\noalign{\hrule}
\noalign{\vskip 5 pt}
Initial Mass & Remnant Mass & Comments \cr
($M_{\odot}$ )& ($M_{\odot}$) & \cr
\noalign{\vskip 5 pt}
\noalign{\hrule}
\noalign{\vskip 5 pt}
$<4.7$ & $0.58 +0.22\times (m/M_\odot-1)$ &White dwarf\cr
[4.7,8.0]& 0.0 & No remnant \cr
[8.0,15.0]& 1.4 & Neutron star \cr
\noalign{\vskip 5 pt}
\noalign{\hrule}
}
}$$
\vfill\eject
\centerline{Table 3. Results of Survey}
$$\vbox{
\halign{\hfil#\hfil&\quad\hfil#\hfil&\quad\hfil#\hfil&\quad\quad\quad#
\hfil&\quad\hfil#\hfil&\quad\hfil#\hfil\cr
\noalign{\hrule}
\noalign{\vskip 2 pt}
\noalign{\hrule}
\noalign{\vskip 5 pt}
$W_0$& $\alpha$ & $U_{\rm t}$ & Lifetime&CW&$A_{min}$\cr
& & ($yr.$) &($yr.$)&($yr.$)\cr
\noalign{\vskip 5 pt}
\noalign{\hrule}
\noalign{\vskip 5 pt}
1 & $-1.5$ & $2.37\times 10^6$ & $32(7.6\times 10^7)$ & $9.2\times 10^6$
&$-0.9$\cr 1 & $-2.5$ & $2.56\times 10^6$ & $108(2.9\times 10^8)$ &
$3.4\times 10^7$&$-0.6$ \cr 1 & $-3.5$ & $2.65\times 10^6$ &
$830(2.2\times 10^9)$ & $2.5\times 10^9$&$-0.8$ \cr 3 & $-1.5$ &
$1.76\times 10^6$ & $60(1.1\times 10^8)$ & $1.4\times 10^7$ &$-0.5$ \cr
3 & $-2.5$ & $1.90\times 10^6$ & $480(9.1\times 10^8)$ & $2.8\times
10^8$&$-0.75$ \cr 3 & $-3.5$ & $1.96\times 10^6$ & S & C &$-0.3$\cr 5 &
$-1.5$ & $1.07\times 10^6$ & $220(2.4\times 10^8)$ & $----$ &$-0.75$\cr
5 & $-2.5$ & $1.16\times 10^6$ & S & $----$ &$-0.3$\cr 5 & $-3.5$ &
$1.19\times 10^6$ & S & $----$ &$-0.15$\cr 7 & $-1.5$ & $5.28\times
10^5$ & S & $1.0\times 10^9$ &$-0.35$\cr 7 & $-2.5$ & $5.71\times 10^5$
& S & C &$-0.1$\cr 7 & $-3.5$ & $5.91\times 10^5$ & S & C &$-0.05$\cr
\noalign{\vskip 5 pt}
\noalign{\hrule}
}
}$$
\bigskip
\centerline{Table 4. Lifetime for different galactocentric distances}
$$\vbox{
\halign{\hfil#\hfil&\quad\hfil#\hfil&\quad\hfil#\hfil&\quad\quad\quad#
\hfil&\quad\hfil#\hfil\cr
\noalign{\hrule}
\noalign{\vskip 2 pt}
\noalign{\hrule}
\noalign{\vskip 5 pt}
$R_{\rm g}$& $U_{\rm t}$ & Lifetime &CW &CW\cr
($Kpc$)	   & 	($yr.$)	 & ($yr.$)  &Family&Lifetime\cr
\noalign{\vskip 5 pt}
\noalign{\hrule}
\noalign{\vskip 5 pt}
4.0 & $1.9\times 10^6$ & $480(9.1\times 10^8)$ & 1 & $2.8\times 10^8$ \cr
10.6 & $5.0\times 10^6$ & $320(1.6\times 10^9)$ & 2 & $2.9\times 10^8$ \cr
18.0 & $8.5\times 10^6$ & $250(2.1\times 10^9)$ & 3 & $2.9\times 10^8$ \cr
47.4 & $2.2\times 10^7$ & $170(3.8\times 10^9)$ & 4 & $2.9\times 10^8$ \cr
\noalign{\vskip 5 pt}
\noalign{\hrule}
}
}$$
\vfill\eject

\parindent 0pt

{\bf Figure Captions}
\medskip
Fig.1  Evolution of the mass with time.  The title of each graph gives
the initial dimensionless central potential and the slope of the mass
function.  The abscissa is time, in standard $N$-body units.  The thin
line denotes the mass of all stars, and the thick line the mass of
those within the tidal radius.

Fig.2 Evolution of Lagrangian radii for the model with initial
dimensionless central potential $W_0 = 3$ and mass function slope
$\alpha = -1.5$.  The thin lines give the radii of spheres, centred on
the density centre, containing corresponding fractions (stated as
percentages) of the total initial mass of all stars. The thick curve
gives the tidal radius, and the abscissa is time in $N$-body units.

Fig.3 As Fig.2 for an intermediate mass function slope.

Fig.4 As Fig.2 for a steep initial mass function.

Fig.5 Evolution of the mass of stars within the tidal radius for fixed
initial parameters $W_0$, $\alpha$ and total mass, but for different
galactocentric radii.  The conversion of the abscissa to astrophysical
time units is given in Table 4.

Fig.6 Dependence of the evolution on the number, $N$, of particles
used in the simulation, for a case with a relatively flat initial mass
function. The ordinate is the mass of stars within the tidal radius.

Fig.7 As Fig.6, but for an intermediate slope of the initial mass
function.

Fig.8 As Fig.6, but for a steep mass function.

Fig.9 Total anisotropy of stars within the tidal radius, as a function
of time in $N$-body units.  The anisotropy parameter $A$ is defined in
eq.(19), and the graphs are arranged as in Fig.1.

Fig.10 Spin angular velocity of one case as a function of $N$-body
time.  The angular velocity $\Omega_c$ is defined in terms of the
component of the spin angular momentum orthogonal to the plane of
motion of the cluster about the galaxy.

Fig.11 The virial ratio, defined by eq.(21), for several
representative cases, plotted against $N$-body time.  Each case is
labelled by initial values of the parameters $W_0$ and $\alpha$.

Fig.12 Form factors $\mu$ and $\nu$ for the contributions to the
potential energy from the gravitational field of the cluster and
galaxy, respectively.  They are defined in terms of the potential
energies, mass and half-mass radius by eqs.(25) and (26).

Fig.13 Theoretical interpretation of tidal disruption.  Definitions:
$r_t$, tidal radius; $E$, total energy of the cluster; $V_s$,
contribution of spin angular momentum to the virial balance (cf.
eq.(7)); $M$, total mass within $r_t$; $r_h$, half-mass radius.  The
thick line plots eq.(28) (a version of the equation of virial balance)
for the stated values of the form factors $\mu$ and $\nu$, and the
thin lines sketch the two contributions (due to the cluster and the
tide) to the right side of eq.(28).  The stars denote the evolution of
a model with stated parameters $W_0$ and $\alpha$, some of which are
labelled with $N$-body time.

\bye